\documentclass[11pt,a4paper]{article}
\begin{document}

\title{ The stable problem in the Rindler space-time  \footnote{ E-mail of Tian:
 hua2007@126.com, tgh-2000@263.net}}
\author{Tian Gui-hua,\ \ Shi-kun Wang,  \ \ Shuquan Zhong\\
School of Science, Beijing University \\
of Posts And Telecommunications, Beijing 100876, P.R. China.
\\Academy of Mathematics and Systems Science,\\ Chinese Academy of
Sciences,(CAS) Beijing 100080, P.R. China.}
\date{}
\maketitle

\begin{abstract}
We carefully study the stable problem of the Rindler space time by
the scalar wave perturbation. Using the two different coordinate
systems, the scalar wave equation is investigated. The results are
different in these two cases. They are analyzed and compared in
detail. The conclusions are: (a) the Rindler space time as a whole
is not stable; (b) the Rindler space time could exist stably only
as a part of the Minkowski space time, and the Minkowski space
time could be a real entity independently; (c) there are some
defects for the scalar wave equation written by the Rindler
coordinates, and it is unsuitable for investigation of the stable
properties of the Rindler space time. All these results might shed
some lights on the  stable properties of the Schwarzschild black
hole. It is natural and not unreasonable for one to infer that:
(a) perhaps the Regge-Wheeler equation is not sufficient to decide
the stable properties; (b) the Schwarzschild black hole as a whole
might be really unstable; (c) the Kruskal space time is stable and
can exist as a real physical entity ; whereas the Schwarzschild
black hole could occur only as part of the Kruskal space time.

\textbf{PACS}: 0420-q, 04.07.Bw, 97.60.-s
\end{abstract}
\section*{I. Motivation}

The stability of the Schwarzschild black hole is a vitally
important problem in general relativity. In 1957, Regge and
Wheeler first treated the problem by the Schwarzschild coordinates
\cite{rw}. They obtained the perturbation equation, that is, the
well-known Regge-wheeler equation. The horizon and the spatial
infinity consist of the boundaries for the perturbation problem.
It is easy to treat the boundary of the spatial infinity; Whereas
the boundary conditions at the horizon are complicated. The
background metric has an apparent singularity at the horizon
$r=2m$ in the Schwarzschild coordinates; this in turn may result
in spurious divergence in the perturbation and cause confusion. A
 proposition was given by Vishveshwara in 1970.  Because the
metric is regular at $r=2m$ in the Kruskal coordinates system, he
transformed the perturbation fields into it. By using the
Schwarzschild time $t=0$ to define the initial time, it is
obtained that the black hole is stable \cite{vish}.

In practice, there is still another problem concerned with the
Schwarzschild coordinates. Due to its metric component $g_{00}=0$
at the horizon $r=2m$, the Schwarzschild time coordinate $t$ loses
its meaning at the horizon. This is the drawback of the
Schwarzschild time coordinate $t$. For example, a particle near
the horizon $r=2m$ falls into the black hole by finite proper
time; but it is described by the Schwarzschild time coordinate as
an infinity process as $t\rightarrow +\infty ,\ r\rightarrow 2m $.
This example shows that the Schwarzschild time coordinate $t$ is a
tortoise coordinate. It is well-known that
$r_*=r+2m\ln(\frac{2m}r-1)$ is a tortoise coordinate.

Actually, this tortoise property of $t$ appears more easily to
escape one's notice, and the researchers had hardly considered its
influence on the stable problem\cite{rw},\cite{vish}: $t=0$ is
taken it for granted as the initial time at the horizon. One
should doubt its validity. We have first taken notice of it and
worked over it. For simplification, one might select some "good"
time coordinate to define the initial time. Using the new "good"
coordinate time, the stable problem should be reinvestigated
again. The Kruskal time $T$ and the Painlev\'{e} time are all
"good" coordinate times. We have employed them to study the
problem and obtained unusual results. By using the Kruskal time
coordinate $T=C=const$ to define the initial time, we found that
its stable properties controversially depend on the sign of the
initial time $T=C=const$: the Schwarzschild black hole is stable
when $T=C \ge 0$, whereas the Schwarzschild black hole is unstable
when $T=C \le 0$\cite{tian1}-\cite{tian4-1}. The same is also true
for the Painlev\'{e} time: when the initial time slice intersects
the future horizon, it is stable; whereas it is unstable as the
initial time slice intersects the past horizon. These unexpected
results are in contrast with the conclusion taken it for granted
that the Schwarzschild black hole is stable.

Which one is  correct? Whether is the Schwarzschild black hole
stable or not? Where and why do the contradictories come from?  Is
it just an effect caused by the tortoise property of the time
coordinate $t$? Actually, we have already replaced it by "good"
time, this by turn excludes this possibility. Perhaps this
complicated situation really shows that the Regge-Wheeler
equation, which was obtained in the Schwarzschild coordinates, is
not sufficient to decide the stable properties. As shown before,
the Schwarzschild coordinate $t$ is a tortoise coordinate as the
well-known tortoise coordinate $r_*$; they do not cover the event
horizon. So, they  are possibly not qualified for the stable
study.

The Kruskal space time is the extension of the Schwarzschild black
hole. The Kruskal coordinates are qualified for study of the
perturbation equation and the stable problem. But the background
metric is varying with time in the Kruskal coordinates
\cite{vish}, subsequently, it is almost impossible to restudy the
problem in the whole Kruskal coordinates. So there is still not a
convincing answer to the stable problem of the Schwarzschild black
hole.

On the ground of the difficulties, we want to study the stable
problem of the Schwarzschild black hole indirectly and are led to
study that of the Rindler space time.

The Rindler space time is one part of the Minkowski space time
whose constant spatial coordinates describe   accelerated
observers with constant accelerations. The interest in Rindler
space time lies in almost the same geometrical structure with that
of the Schwarzschild black-hole. Actually, the Hawking radiation
in the Schwarzschild black-hole is closely connected with the
Unruh effect in the Rindler space time\cite{unru},\cite{boul}.

The Rindler coordinates are the same as those of the Schwarzschild
time coordinates; its time is a tortoise coordinate too. What
effect would it have on the stable study of the Rindler space
time? In order to find it, we will study the Klein-Gordon equation
in the whole Rindler coordinates in section II and discuss the
stable properties of the Rindler space time. The same problem is
treated in section III by the Minkowski coordinates; some
arguments are given and conclusion is obtained in section IV.

\section*{II.The scalar field equation in the Rindler space time}

The metric of the Rindler space time is
\begin{equation}
ds^{2}=-z^2dt^{2}+dz^{2}+dx^{2}+ dy^{2}\label{orimetric-r}
\end{equation}
with $0<z<+\infty $, $t,\ x,\ y\in (-\infty,\ +\infty)$. The
Minkowski space time's metric is
\begin{equation}
ds^{2}=-dT^{2}+dZ^{2}+dx^{2}+ dy^{2}.\label{orimetric-m}
\end{equation}
The transformation equations between them are
\begin{equation}
z=\sqrt{Z^2-T^2}\label{relation zZ1}
\end{equation}
and
\begin{equation}
e^{2t}=\frac {Z+T}{Z-T},\label{relation zZ0}
\end{equation}
or
\begin{equation}
Z=z\cosh t\label{relation Zz1}
\end{equation}
and
\begin{equation}
T=z\sinh t.\label{relation Zz0}
\end{equation}
Of course, the Rindler space time's completion is the Minkowski
space time, and corresponds to only the part $Z>0,\ Z^2-T^2>0$ of
the Minkowski space time. $Z^2-T^2=0$ is the horizon of the
Rindler space time. The horizon could be denoted alternatively by
$z=0$ in the Rindler coordinates. The metric (\ref{orimetric-r})
is obviously singular at the horizon.

In the Rindler coordinates, the Klein-Gordon equation of the
scalar field with mass $\mu$ is
\begin{equation}
\frac 1 {\sqrt{-g}} \frac {\partial }{\partial
x^{\mu}}\left[\sqrt{-g}g^{\mu \nu}\frac {\partial \Psi}{\partial
x^{\nu}}\right]-\mu^2\Psi =0, \label{kg eqwuation}
\end{equation}
which turns out as
\begin{equation}
-\frac 1 {z}\frac{\partial^{2}\Psi}{\partial t^{2}}+\frac 1 {z}
\frac {\partial }{\partial z}\left[z\frac {\partial \Psi}{\partial
z}\right]+ \frac{\partial^{2}\Psi}{\partial
x^{2}}+\frac{\partial^{2}\Psi}{\partial
y^{2}}-\mu^{2}\Psi=0.\label{kg in r}
\end{equation}

The coordinates $t,\ x,\ y$ range from $-\infty $ to $+\infty $,
therefore, the normal mode decomposition of the scalar field
$\Psi$ must be
\begin{equation}
\Psi =\psi(z) e^{-i\omega t+ik_1x+ik_2y}
\end{equation}
with $\psi$ satisfying the following equation
\begin{equation}
 \frac {d}{d z}\left[z\frac {d\psi}{d z}\right]-\frac {\omega
 ^2}z\psi+
z\left[\mu^{2}+k_1^2+k_2^2\right]\psi=0.\label{kg2 in r}
\end{equation}
The equation (\ref{kg2 in r}) is Bessel's equation with its order
being $\pm i\omega $\cite{boul}. Its solutions are the modified
Bessel function $I_{\pm i\omega }(\sqrt{\mu^{2}+k_1^2+k_2^2}\
z)$.\footnote{when $i\omega $ is an positive integer, $I_{-
i\omega }$ is replaced by $K_{- i\omega }$.} Suppose the frequency
$\omega =\alpha +i\beta $ with $\alpha , \ \beta $ real and $\beta
>0$, then $I_{- i\omega }(\sqrt{\mu^{2}+k_1^2+k_2^2}\ z)$ goes to
infinity exponentially as $z\rightarrow \infty $ and vanishes at
$z=0$. On the contrary, $I_{+ i\omega
}(\sqrt{\mu^{2}+k_1^2+k_2^2}\ z)$ goes to infinity at $z=0$ and
falls off to zero exponentially at $z\rightarrow \infty
$\cite{boul}.

The  boundaries of the  equation (\ref{kg2 in r}) consist of
$z=0$, the horizon $H^ \pm $ of the Rindler space time, and
$z\rightarrow \infty $ of the infinity. We must demand the scalar
field function $\Psi$ initially well-behaved at the boundaries.
For the infinity, we could require the field $\Psi$ falling off to
zero initially. So we just select $I_{i\omega
}(\sqrt{\mu^{2}+k_1^2+k_2^2}\ z)$ to satisfying the infinity
boundary condition initially with the frequency whose imaginary is
positive, that is,
\begin{equation}
\Psi = e^{-i\omega t+ik_1x+ik_2y}I_{ i\omega
}(\sqrt{\mu^{2}+k_1^2+k_2^2}\ z)\label{I1 in rindler}.
\end{equation}
Generally, researchers would use the Rindler coordinate time $t=0$
as the initial time and study the problem. By Eq.(\ref{I1 in
rindler}), the scalar field is regular at the infinity
($z\rightarrow \infty $)at $t=0$. But  $I_{ i\omega
}(\sqrt{\mu^{2}+k_1^2+k_2^2}\ z)$ blows up at the horizon $z=0$ at
$t=0$, that is, $\Psi $ is initially blows up at the horizon for
the positive imaginary of the frequency ($\beta
>0$). This shows that $\beta >0$ is unacceptable for the scalar
field because it behaves badly at the initial time $t=0$.
Therefore the scalar field is stable and the Rindler space time is
also stable to some extent ( confining to the perturbation of the
scalar field ).

But the metric is singular at the horizon $z=0$, and the time
coordinate $t$ even loses its meaning there due to the fact
$g_{00}|_{z=0}=0$. In reality, the past and future horizons
correspond to $t\rightarrow -\infty $ ,$t\rightarrow +\infty $
respectively. Their intersection is $T=Z=0$ in Minkowski
coordinates, but it corresponds to $z=0$ and $t$ from $-\infty$ to
$+\infty$ in the Rindler coordinates. This example clearly shows
the Rindler time coordinate $t$ is meaningless and $t=0$ should
not be used as initial time containing the horizon. Therefore, we
can not be sure that the above stable conclusion is really right;
we should also employ the Minkowski time coordinate to treat the
problem again.

Now, we make use of the Minkowski time $T$  to define the initial
time. By the transformation equations (\ref{relation
zZ1}),(\ref{relation zZ0}), we could rewrite the scalar field
$\Psi$ as the function of the independent variables $T,\ Z,\ x,\
y$ as:
\begin{equation}
\Psi =A e^{\frac{-i\omega
t}2\ln{\frac{Z+T}{Z-T}}}e^{ik_1x+ik_2y}I_{ i\omega
}(\sqrt{\mu^{2}+k_1^2+k_2^2}\sqrt{Z^2-T^2})\label{I2 in
Minkowski}.
\end{equation}
Excluding the horizon, the scalar field $\Psi$ is bounded at the
initial time $T=C=const$ for the positive frequency $\Im \omega
=\beta >0$. At the horizon $Z^2-T^2=0$, the asymptotic form of the
scalar field $\Psi$ in Eq.(\ref{I2 in Minkowski}) is
\begin{equation}
\Psi \propto A \left({\frac{Z+T}{Z-T}}\right)^{\frac{-i\omega
}2}\left(Z^2-T^2\right)^{\frac{i\omega }2}=A
\left(Z-T\right)^{i\omega } \label{I3 in Minkowski}.
\end{equation}
Whether or not the equation (\ref{I3 in Minkowski}) is initially
divergent at the horizon really depends on the sign of the initial
time $T=C$. When we select the initial time $T=C\ge 0$, The
initial time slice $T=C\ge 0$ intersects the future horizon that
corresponds to $Z=T=C$. So, the scalar field $\Psi$ goes to
infinity initially at $Z=T=C$ (the horizon). This excludes the
possibility of the positive imaginary frequency, and subsequently
ensures its stability and the Rindler space time respectively to
some extent.

Alternatively, we could also select the initial time $T=C<0$. The
initial time slice $T=C$ intersects the past horizon that is
denoted by $Z=-T=-C$, so the scalar field $\Psi$ is also bounded
initially at the horizon for the positive imaginary frequency
$\beta >0$. In this very case, the scalar field $\Psi$ is not
stable and the Rindler space time consequently is also unstable
with respect to the scalar field perturbation.

Here comes some discussion.

It has been shown that the stable properties of the Rindler space
time vitally depend on using what time coordinate to define the
initial time. If we use the Rindler time, the scalar field is
stable. On the ground of that, the Rindler time coordinate loses
its time meaning in the horizon, one should choose "good" time to
define the initial time. Of course, the Minkowski time is one of
the "good" time. When we choose the Minkowski time to define the
initial time, the scalar field' stable properties become very
intricate and depend on where the initial time slice intersects
the horizon. The Rindler space time is stable when the initial
time slice intersects the future horizon; while it is unstable as
the initial time slice intersects the past horizon. The
controversial conclusions  is almost the same  as  those for the
Schwarzschild black hole \cite{tian1}-\cite{tian4-1}. This problem
is unsolved in the Schwarzschild black hole. It is suggested that
one should study the stable problem of the Schwarzschild black
hole in the whole Kruskal coordinate system to get a definite
answer to the controversy\cite{vish}. But it becomes almost
impossible because the background metric in the Kruskal coordinate
system is not even stationary. So, one has not obtained an
absolutely and convincingly unambiguous answer to the stable
problem of the Schwarzschild black hole.

The results are similar in the two cases. This is a truly good
news. We could really get something interesting for the
Schwarzschild black-hole from the present results. The Minkowski
space time is the extension of the Rindler space time, and is
luckily static in nature. This static property of the Minkowski
spcae time presents a striking contrast to that of  the whole
Kruskal space time, i.e., the extension of the Schwarzschild black
hole. We could easily investigate the perturbation equation of the
Rindler space time in whole Minkowski coordinates. It is not
difficult to obtain the result. It might give some clues to the
stable problem of the Schwarzschild black hole.

\section*{III. The scalar field equation in the Minkowski space time}

The scalar field equation in the Minkowski space time is
\begin{equation}
-\frac{\partial^{2}\Psi}{\partial T^{2}}+ \frac {\partial ^2
\Psi}{\partial Z^2}+ \frac{\partial^{2}\Psi}{\partial
x^{2}}+\frac{\partial^{2}\Psi}{\partial
y^{2}}-m^{2}\Psi=0.\label{kg in m}
\end{equation}
The Rindler space time corresponds the region where the
coordinates $T,\ x,\ y$ range from $-\infty $ to $+\infty $ and
$Z>0,\ Z^2-T^2>0$, therefore, the normal mode decomposition of the
scalar field $\Psi$ must be
\begin{equation}
\Psi = Ae^{-i\omega T+ik_3Z+ik_1x+ik_2y}\label{kg1 in m}
\end{equation}
where $k_3$ must satisfy the following equation
\begin{equation}
k_3=\pm \sqrt{\omega^2-k_1^2-k_2^2-m^2}.
\end{equation}
In order to make the scalar field $\Psi$ initially finite over
$-\infty<x,\ y<+\infty $, the numbers $k_1,\ k_2$ of course are
real. If we similarly choose the frequency $\omega =i\beta $ with
$\beta
>0$, the above equation becomes
\begin{equation}
k_3=\pm i\sqrt{\beta^2+k_1^2+k_2^2+m^2}.
\end{equation}
We could select $k_3=i\gamma $ with
$\gamma=\sqrt{\beta^2+k_1^2+k_2^2+m^2}>0$, then the scalar field
$\Psi$ is
\begin{equation}
\Psi =Ae^{\beta T}e^{-\gamma Z+ik_1x+ik_2y}.\label{psi in M}
\end{equation}
The boundaries of the Rindler space time consist of the infinity
$Z\rightarrow +\infty$ and the horizon $Z^2-T^2=0$. Because the
Minkowski metric is regular everywhere including the horizon, we
could directly use the Minkowski time $T=0$ to define the initial
time. By Eq.(\ref{psi in M}), we could see the scalar field $\Psi$
is well-behaved everywhere corresponding to the Rindler space time
at the initial time $T=0$. But it blows up to infinity as the time
$T$ goes to infinity for $\beta >0$.

Directly, it is easy to see that  the Rindler space time as a
whole is unstable with respect to the scalar field perturbation.

\section*{IV. Arguments and conclusion}

We just obtain the conclusive instability of the whole Rindler
space time by the Minkowski coordinates. We now check the study in
the Rindler coordinates in section II. The result definitely
supports the view that the Rindler time coordinate $t$ is not
suitable for one to study the stable problem containing the
horizon where $t$ loses its meaning. This might suggest that the
Schwarzschild time $t=0$ should not be qualified as initial time
for study of the Regge-Wheeler equation in the Schwarzschild black
hole.

The result also supports that only the instability in section II
by using the Minkowski time $T=C<0$ is correct. This can be
explained easily. The Minkowski space time consists of four parts:
the right region $R$, i.e.,the Rindler space time $Z>0,\
Z^2-T^2>0$; the left region $L$, i.e.,$Z<0,\ Z^2-T^2>0$; the past
region $P$,i.e., $T<0,\ Z^2-T^2<0$; the future region $F$,i.e.,
$T>0,\ Z^2-T^2<0$. At the initial time slice $T=C<0$, the Rindler
space time could be influenced by the part $P$ where every
particle must leave and come into either the Rindler space time or
the other parts $L$ and $F$. In this way, the Rindler space time
is unstable. Similarly, when the initial time slice intersects the
past horizon of the Schwarzschild black hole, it it easy to see
that its instability comes from the influence of the white hole.
Though the Kruskal metric of the Schwarzschild black hole is not
stationary, and one has not obtained the same conclusive result as
that for the  Rindler space time in the Minkowski coordinates; it
is nevertheless not unreasonable to infer that the Schwarzschild
black hole as a whole might really be unstable in comparison with
the case in Rindler space time. Of course, one must go further to
get the conclusion definitely.

Why is the stable conclusion wrong when using the Minkowski time
$T=C>0$? Furthermore, it is well-known that the Minkowski is
stable, how does one harmonize  our instability conclusion of the
Rindler space time with this fact? Now we once again investigate
the process for our conclusion. By the Eq.(\ref{psi in M}), only
when the scalar field $\Psi$ is completely confined to the Rindler
space time,  could it has positive imaginary frequency. This by
turn shows that the Rindler space time as a whole is unstable. On
the contrary, when the whole Minkowski space time is served as its
background ($Z$ ranges from $-\infty$ to $+\infty$ ), the
Eq.(\ref{psi in M}) clearly excludes the possibility of the
positive imaginary frequency of the scalar field $\Psi$.
Therefore, the Minkowski space time is definitely stable.
Actually, this is just the well-know fact of the Minkowski space
time; no new conclusion comes. Really, there lies some
inconsistency or contradiction in these conclusions. How to
explain them?  We give some naive arguments below.

When  the Minkowski time $T=C>0$ is the initial time, only the
part $F$ borders upon the Rindler space time initially.
Classically, the part $F$ could not influence the Rindler space
time, but this might be false from the quantum theory. It is
well-known  from the de-Broglie's theory of matter wave that the
phase velocity of the relativistic particles' might be possibly
greater than that of the light. The scalar wave equation could
also be regarded as a quantum wave equation for a freely moving
particle with relativistic velocity. From the wave eqs.(\ref{kg2
in r})-(\ref{I3 in Minkowski}) and by the WKB method, it is not
difficult to see that the phase velocity of the wave is not
greater than that the light velocity at the horizon. So, the part
$F$ could not have any effect on the Rindler space time in quantum
way (or by wave theory). This insures the stable conclusion. But
now we know the conclusion is wrong, there must exist something
wrong in the wave equation (\ref{kg2 in r}) or (\ref{kg in r}). To
see this clearly, let's inspect the wave equations (\ref{kg in m})
and (\ref{kg1 in m}) in the Minkowski coordinates. The
Eq.(\ref{kg1 in m}) presents a striking contrast to Eq.(\ref{kg2
in r}), its phase velocity can easily greater than the light
velocity at the horizon. Therefore the part $F$ has definitely
influence on the Rindler space time, so does the part $L$. The
Rindler space time as a whole is impossible to be stable, only the
whole Minkowski space time is stable. From the above argument, we
see that the wave equation (\ref{kg in r}) loses some useful
information at the horizon and might result in error in the stable
study for the Rindler space time.

Therefore, we reinforce our conclusion again: the whole Minkowski
space time is a real entity, while the Rindler space time can not
exists independently. Actually, the Rindler space time cannot
exist if the left region $L$ does not exist.

When compared with the stability study of the Schwarzschild black
hole, one could reasonably infer the following: (1) the
perturbation equation should be in the Kruskal space time, though
it may be difficult; (2)the Regge-Wheeler equation is possibly not
sufficient to decide the stable properties, and it surely loses
some useful information at the horizon, which will be our further
study; (3) if the Schwarzschild black hole exists stably, the
other region not to communicate with it must exist, that is, the
Kruskal space time is a real physical entity.

\section*{Acknowledgments}

We are supported in part by the National Key Basic Research and
Development Programme of China under Grant No. 2004CB318000; the
National Natural Science Foundations of China under Grant
No.10475013, No.10375087, No.10373003
 and the post-doctor foundation of China.

\end{document}